# In Vivo Study of Bone Growth Around Additively Manufactured Implants with Ti-6Al-4V and Bioactive Glass Powder Composites

Chih-Yu Lee[1], Pei-Ching Kung[2], Chih-Chieh Huang[2], Shao-Ju Shih[3], E-Wen Huang[2], San-Yuan Chen[2], Meng-Huang Wu[4-5]*, Nien-Ti Tsou[2]*

[1] Department of Materials Science and Engineering, University of Maryland, College Park, MD 20742, USA

[2] Department of Materials Science and Engineering, National Yang Ming Chiao Tung University, Hsinchu, Taiwan (ROC)

[3] Department of Materials Science and Engineering, National Taiwan University of Science and Technology, No. 43, Sec. 4, Keelung Rd., Taipei 10607, Taiwan

[4] Department of Orthopaedics, School of Medicine, College of Medicine, Taipei Medical University, No. 250, Wuxing St., Xinyi District, Taipei 11031, Taiwan

[5] Department of Orthopedics, Taipei Medical University Hospital, No. 252, Wuxing St., Xinyi District, Taipei 11031, Taiwan* Corresponding authors

Abstract

Osseointegration is crucial to the success of biomedical implants. Additive manufacturing of implants offers a high degree of design freedom, enabling precise control over implant geometry and material composition. Bioactive glass (BG) can substantially enhance bone binding and bioactivity; however, limited research has been conducted on its incorporation into additively manufactured implants. The performance of BG varies depending on the incorporation method, and the spatial and temporal evolution of its integration remains unclear. In this study, we synthesized Ti-6Al-4V/58S BG composites by using the selective laser melting method and systematically compared the effects of BG coating and doping in additively manufactured implants. In vivo histological results from animal tests were statistically analyzed and discussed in terms of osseointegration over 4- and 12-week periods. Bone-to-implant contact (BIC) and bone density (BD) were used as quantitative metrics to evaluate interactions between the implants and surrounding bone. Our findings indicate that both BG-doped and BG-coated implants accelerated bone ingrowth during the early stages of healing. BG-coated implants demonstrated a greater improvement than did pure 3D-printed Ti-6Al-4V implants. However, the effects of BG became nonsignificant during the later healing stage (12 weeks). This study provides a foundation for systematically investigating BG incorporation methods in 3D-printed biomedical implants and their effect on osseointegration.

1. Introduction

Implant materials play a pivotal role in the success of orthopedic and dental surgeries[1]. Among these materials, titanium alloys, especially Ti-6Al-4V, are widely known for their excellent biocompatibility, corrosion resistance, and mechanical strength[2–4]. However, the biomechanical mismatch between titanium implants and surrounding tissues remains a substantial problem[5]. Several studies have indicated that differences in stiffness and elastic modulus lead to stress-shielding, which impairs bone healing and remodeling[6, 7]. Therefore, researchers have begun to focus on developing bone-like materials through structural modifications[8] and compositional adjustments[9] to mitigate stress-shielding and promote osseointegration.

Bioactive glasses (BGs) and BG composite materials, which were first introduced by Hench et al.[10, 11], have attracted considerable attention in tissue engineering and drug delivery[12–15] because of their excellent bioactivity, degradability, and low inflammatory response. BGs form hydroxyapatite (HA) layers on implants and release ion dissolution products, establishing a robust interface between hard and soft tissues and promoting bone growth[16–18]. Therefore, extensive research has been conducted on BG-coated layers[19–22], with a focus on enhancing implant performance. Incorporating BG composites into pure metallic biomaterials could improve bone binding and bioactivity. However, the granularity of BGs limits their reliability as space-making devices. In addition, BGs lack osteoinductive properties and cannot induce bone formation at ectopic sites.[23]

Various types of BGs have been developed, including 58S BGs, which are able to promote apatite layer formation, facilitate rapid bone bonding, support degradation and resorption, and induce osteoblastic differentiation[24]. In particular, sol-gel-derived 58S BGs are considered a promising alternative to glass-melt-derived 45S5 BGs because of their superior capacity to induce osteoblastic differentiation[25]. Fathi et al.[26,27] have demonstrated that 58S BG-coated 316L stainless steel improved biocompatibility and osteointegration, leading to earlier implant stabilization and decreased healing time. Our previous study[28] indicated that the spray-drying method for preparing 58S BGs reduced contamination and improve production efficiency compared with the conventional glass-melting and sol-gel methods. In addition, 58S BGs exhibit better bioactivity and a faster HA growth rate than BGs with other compositions do[29]. Thus, 58S BGs are particularly suitable for biomedical applications and were selected for use in the current study.

Additive manufacturing (AM) technology has been widely applied to both ceramics and alloys to achieve complex geometries[8, 15, 30–32] and has been used to produce BG composite materials. Lam et al.[28] performed histological and histomorphometric analyses to evaluate the properties and in vivo performance of titanium (Ti)-based alloy and BG composite materials fabricated using the selective laser melting (SLM) method. However, detailed studies are required to better understand bone healing around BG-coated and BG-doped implants, particularly the chemical and physical interactions of BGs and their temporal and spatial evolution in clinical applications. In this study, we investigated the effect of 0.25 wt% 58S BG applied as both a coating and a dopant in Ti-6Al-4V implants. Animal implantation tests were conducted using rabbits. Histological results were analyzed using one-way analysis of variance (ANOVA) to assess the osteointegration over short and long periods. Quantitative parameters, including bone-to-implant contact (BIC) and bone density (BD), were used to evaluate interactions between the implants and surrounding bone. This study demonstrates the synthesis of high-performance Ti-6Al-4V/BG composites and provides a comparative analysis of the efficacy of BG doping versus coating approaches.

2. Materials and Methods

2.1.1 Preparation of Implants

In this study, three types of implants were evaluated: Implant I (pure Ti-6Al-4V), Implant II (BG-coated Ti-6Al-4V), and Implant III (BG-doped Ti-6Al-4V). All implants were fabricated using the same AM process performed at the Industrial Technology Research Institute (ITRI) in Taiwan. This process involved the use of an SLM machine (ITRI-AM100, Tainan City, Taiwan) with a laser power of 170 W and a scan speed of 1250 mm/s. The implants were designed to replicate the geometry of a commercial Ti-6Al-4V implant, ITI 033.512S (Straumann, Basel,

Switzerland), with a diameter of 3.3 mm and a length of 10 mm. Ti-6Al-4V commercial powder (Titanium Ti64ELI, EOS, Germany) with a particle size of 15–45 μm was used as the base material in this study. In addition, 58S BG powder was selected for its excellent bioactivity.

Implant I was fabricated and designated as the control group to investigate the effect of the compositional addition of BG. Implant II was prepared by immersing Implant I in a BG suspension for 10 minutes, followed by drying in an oven at 70°C for 12 hours. The BG coating suspension was prepared by adding 2.5 g of 58S BG to 95.5 g of deionized (DI) water along with 2.0 g of 95 wt% Type I collagen binder (Horien, Taichung, Taiwan). The mixture was stirred at room temperature for 4 hours. Implant III was manufactured using BG-doped Ti-6Al-4V powder. The Ti-6Al-4V commercial powder was doped with 0.25 wt% 58S BG powder that was synthesized as described in our previous study[29]. Specifically, 0.25 g of 58S BG powder and 999.75 g of Ti-6Al-4V commercial powder were added to 120 mL of DI water and stirred at 100°C for 3 hours. The resulting mixture was dried in an oven at 70°C for 24 hours. The concentration of 0.25 wt% was selected because it enabled the use of a flow time of less than 40 seconds, which was deemed suitable for achieving optimal mechanical properties for 3D printing, as determined in our previous study[28]. The preparation of doped BG is discussed in the following section. Figure 1 provides an overview and scanning electron microscopy images (HITACHI S-3400, Tokyo, Japan) of Implants I and III, demonstrating that the printing quality of the BG-doped Ti-6Al-4V powders was acceptable.

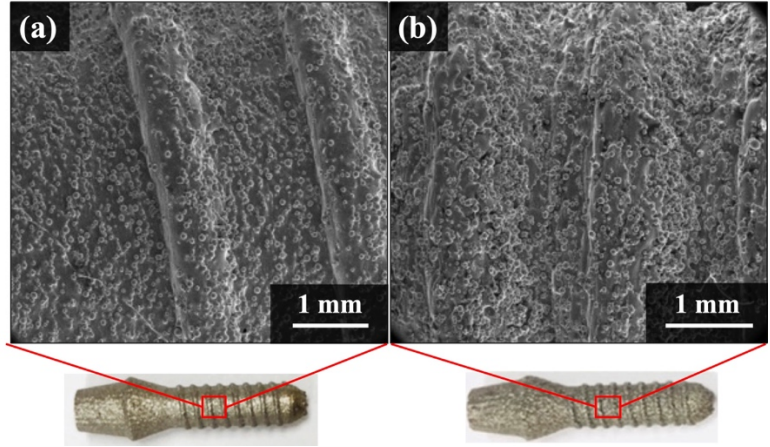

*Figure 1. SEM images of (a) Implants I (pure Ti-6Al-4V) and (b) Implant III (BG-doped Ti-6Al-4V). Implant photos are presented as insets in the corresponding images.*

2.1.2 Preparation of BGs

The BG powder was synthesized through spray pyrolysis by using a 58S composition (60 mol% silicon dioxide, 35 mol% calcium oxide, and 5 mol% phosphorus pentoxide). The solid precursors included 6.70 g of tetraethyl orthosilicate (99.9 wt%, Showa, Japan), 1.40 g of calcium nitrate tetrahydrate (98.5 wt%, Showa, Japan), and 0.73 g of triethyl phosphate (99.0 wt%, Alfa Aesar, USA). These compounds were mixed with 120.00 g of ethanol containing 3.20 g of 0.5 M diluted hydrochloric acid. The resulting solution was stirred at room temperature for 24 hours to ensure homogeneity. The homogeneous solution was then transferred to an ultrasonic atomizer (KT-100A, King Ultrasonic, New Taipei City, Taiwan) operated at a frequency of 1.67 MHz. The atomized droplets were directed into a tube furnace (D110, Dengyng, New Taipei City, Taiwan) with three heating zones set at 250°C, 550°C, and 300°C for preheating, calcination, and cooling, respectively. At the furnace exit, a high voltage of 16 kV was applied to charge the surface of the powders. The charged powders were subsequently neutralized and condensed within a grounded stainless-steel electrostatic collector.

2.2 In Vivo Experiments

2.2.1 Animals

All in vivo experiments were conducted in accordance with the ARRIVE guidelines[33]. The procedures for the care and use of research animals adhered to Taiwanese regulations, ISO 10993-6:2016, and the Good Laboratory Practice for Nonclinical Laboratory Studies (Ministry of Health and Welfare, R.O.C., 3rd ed., 2006). Four male New Zealand rabbits (NZRs) aged between 6 and 7 months (average: 6.5 months) and weighing between 3.2 and 3.6 kg (average: 3.4 kg) were used in this study. The rabbits underwent surgical procedures and were individually housed in cages at the institute (Master Laboratory Co., LTD, Hsinchu, Taiwan) under controlled environmental conditions, including a temperature of 18°C–21°C, natural lighting, moderate moisture, and appropriate air circulation. During the study period, the rabbits were provided with Prolab Rabbit Diet (Lab Diet, PMI Nutrition International, USA) and were given ad libitum access to water.

2.2.2 Surgical Procedures and Animal Sacrifice

The four NZRs were randomly divided into two groups, and each group was sacrificed at either week 4 or 12 following implantation surgeries. A total of eight implants (four Implant I, two Implant II, and two Implant III) were randomly assigned to the left and right femurs of the rabbits in each group. Each femur contained two experimental sites, resulting in four implants per animal. After exposing the femurs, an implanter was used to perforate the experimental sites. The implant diameters were precisely matched to the implant beds in the bone cortex to prevent the ingrowth of fibrous tissue. After careful placement of the implants, the surgical wounds were closed, and the rabbits were administered the antibiotic gentamycin (5 mg/kg intramuscularly) for three consecutive days to prevent infection.

2.2.3 Histological Processing and Statistical Analysis

To evaluate bone healing in peri-implant areas, histological analysis was performed at 4 and 12 weeks postimplantation. Specimens retrieved from the femurs were fixed in 10% neutral formalin for three days at room temperature. These specimens were then dehydrated in a graded ethanol series (60% to 100%) for seven days and embedded in polymethyl methacrylate. The embedded blocks were sectioned into 500-µm-thick slices by using a low-speed precision cutter (IsoMet 11-1280-170, Buehler, IL, USA). The sections were subsequently ground to a thickness of approximately 5 µm, polished, and stained with aniline blue for optical microscopy.

Two parameters were analyzed, namely, (1) BIC and (2) BD, in the region of interest (ROI), which was defined as the area between the implant threads. The ROI is illustrated in Figure 2 as the region between the red dashed line connecting the thread tips and the implant surface. BIC was calculated as the percentage of the total length of the line between the bone (including the newly formed bone, osteoid, and mineralized bone) and the implant relative to the total implant length within the ROI. BD was determined by calculating the percentage of the area occupied by the bone within the ROI. Histomorphometric analysis was conducted using the image postprocessing software ImageJ (U.S. National Institutes of Health, MD, USA).

A parametric ANOVA was conducted to investigate the effects of material composition and implant fabrication on bone healing at weeks 4 and 12. First, a global test was performed using the statistical function "f_oneway()" from the SciPy package[34]. Then, pairwise multiple comparisons between groups were conducted using Student's *t* test from the Scikit-Posthocs package. Significance was defined as $p < 0.05$ ($p < 0.05$*), and $p < 0.001$ was considered highly significant ($p < 0.001$**). The correlation between BIC and BD was analyzed using Spearman rank correlation coefficients.

3. Results

3.1 Descriptive Histological Analysis

The healing of the surgical sites progressed uneventfully and without complications in all NZRs. No signs of inflammation were observed around any of the implants. Figure 2 presents light micrographs depicting bone growth around Implants I to III after 4 and 12 weeks. The blue regions in the images represent the bone, and detailed bone integration within the threads is highlighted in the enlarged images. Overall, all three types of implants exhibited substantial bone growth around the middle portions of the implants after 4 weeks of healing. As illustrated in Figure 2(b) and 2(c), Implants II and III (BG-coated and BG-doped Ti-6Al-4V) promoted greater bone growth than did Implant I (pure Ti-6Al-4V), which led to less new bone tissue in the peri-implant areas [Figure 2(a)]. Enlarged images of the threads at the middle portions of the three implants revealed white regions near the thread surface in Implant I, which are marked by red circles in Figure 2(d). By contrast, improved osseointegration was observed in Implants II and III, as depicted in Figure 2(e) and 2(f), respectively.

After 12 weeks of healing, a significant volume of new bone formation was observed, particularly around the middle and bottom sections of all three types of implants, as illustrated in Figure 2(g), 2(h), and 2(i). However, white regions were present in the newly formed bone areas surrounding Implants II and III [Figure 2(h) and 2(i)]. Thus, the average BD values of Implants II and III were lower than that of Implant I. Enlarged images of selected threads [Figure 2(j), 2(k), and 2(l)] revealed that the surfaces of all three implant types supported effective bone osseointegration.

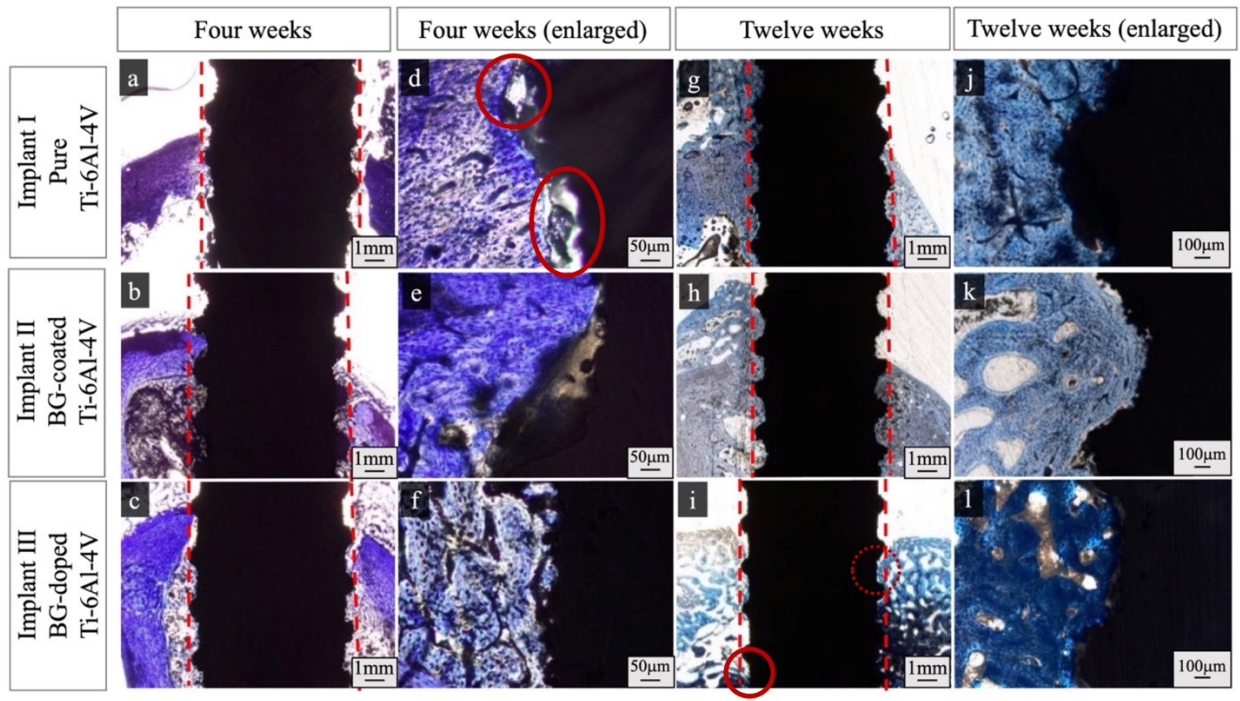

*Figure 2. Light micrographs of bone growth around Implants I to III after (a)–(c) 4 and (g)–(i)12 weeks; detailed bone/tissue integration within the threads is revealed in the enlarged images, (d)–(f) and (j)–(l).*

The numerical histomorphometric results for BD and BIC at weeks 4 and 12 are presented in Table 1. At week 4, Implant II (BG-coated Ti-6Al-4V) had the highest average BD (68.06%) and BIC (67.05%) values, whereas Implant I (pure Ti-6Al-4V) had the lowest BD (15.59%) and BIC (13.58%) values. At week 12, the BD and BIC values for Implant II decreased to 44.79% and 49.56%, respectively, whereas those for Implant I significantly increased to 64.3% and 64.07%, respectively. Additionally, Implant III (BG-doped Ti-6Al-4V) exhibited increased values at week 12 (BD: 58.61%; BIC: 48.22%). The reduction in bone volume around Implant II after the fourth week may be attributable to a marginal bone loss effect, which is discussed further in Section 4.

*TABLE 1 Histomorphometric results of average BIC and BD values at weeks 4 and 12.*

| Time Point | Implant | Average | Average |
| --- | --- | --- | --- |

|  |  | Bone Density (%) | Bone in Contact (%) |
|---|---|---|---|
| Fourth week | I | 15.59 | 13.58 |
|  | II | 68.06 | 67.05 |
|  | III | 47.78 | 40.98 |
| Twelfth week | I | 64.30 | 64.07 |
|  | II | 44.79 | 49.56 |
|  | III | 58.61 | 48.22 |

3.2 Histomorphometric Analysis of BD and BIC

A parametric ANOVA was conducted to evaluate the effects of the test and control materials at weeks 4 and 12. The analysis began with a global test, followed by post hoc comparisons. The BD and BIC results are presented in Figure 3, where $p$ values smaller than 0.001 are considered highly significant and are denoted by **. BD was significantly associated with the biomaterials and the healing period ($p < 0.0001$ in ANOVA). In the post hoc tests, at week 4, both Implant II (BG-coated) and Implant III (BG-doped) had significantly higher BD values than did Implant I (pure Ti-6Al-4V; $p < 0.0001$ and $p < 0.0004$, respectively). However, by week 12, the BD values did not significantly differ among the three implant types ($p = 1$, 0.96, and 0.45, respectively). The BD of Implant I significantly increased between weeks 4 and 12 ($p < 0.0001$).

BIC exhibited a trend similar to that of BD and was significantly associated with the biomaterials and the healing period ($p < 0.0001$ in ANOVA). In the post hoc tests, after 4 weeks of healing, Implant II (BG-coated) had significantly higher BIC values than did Implant I (pure Ti-6Al-4V; $p < 0.0001$). By week 12, the BIC values did not significantly differ among the three implant types ($p = 0.65$, 0.99, and 0.53, respectively). Consistent with the observations for BD, the BIC value of Implant I significantly increased between weeks 4 and 12 ($p < 0.0001$).

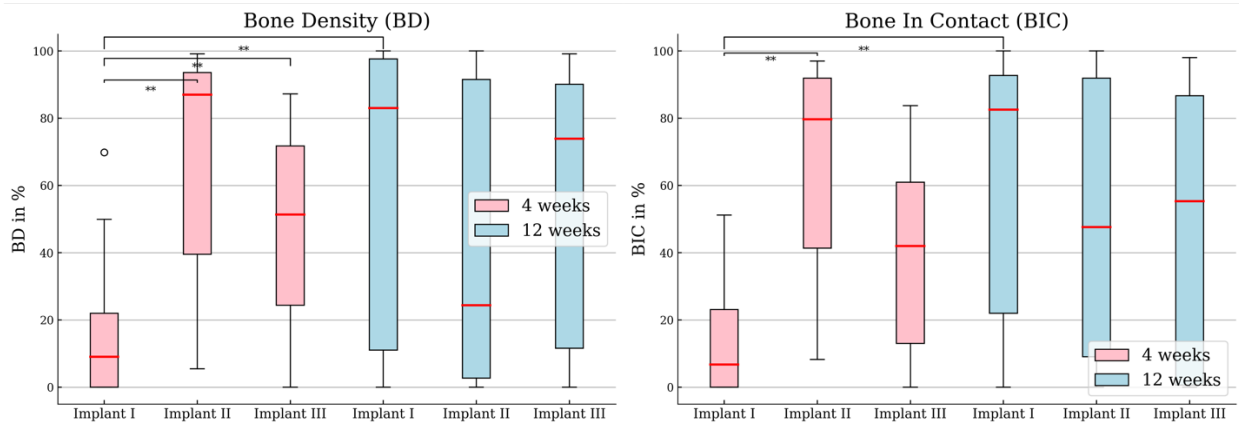

*Figure 3. Histomorphometric results for BD and BIC.*

3.3 Correlation Analysis

A Spearman rank correlation was conducted to evaluate the relationship between BD and BIC at weeks 4 and 12 (Figure 4). At week 4, BD and BIC exhibited strong positive correlations with a high linear dependence (correlation coefficient: +0.96, $p < 0.0001$). However, at week 12, the correlation weakened to a moderate positive relationship (correlation coefficient: +0.64, $p < 0.0001$). This decrease is attributable to data points, primarily corresponding to Implant III, that indicated either "high BIC but low BD" or "high BD but low BIC."

The "high BIC but low BD" scenario indicates the presence of regions where bone adhered well to the implant surface, forming a thin layer, whereas in other peri-implant areas, limited bone growth was noted. An example of this is indicated by a solid red circle in Figure 2(i). Notably, "high BD but low BIC" indicated substantial bone growth in peri-implant areas with minimal contact on the implant surface. This phenomenon is indicated by a dotted red circle in Figure 2(i).

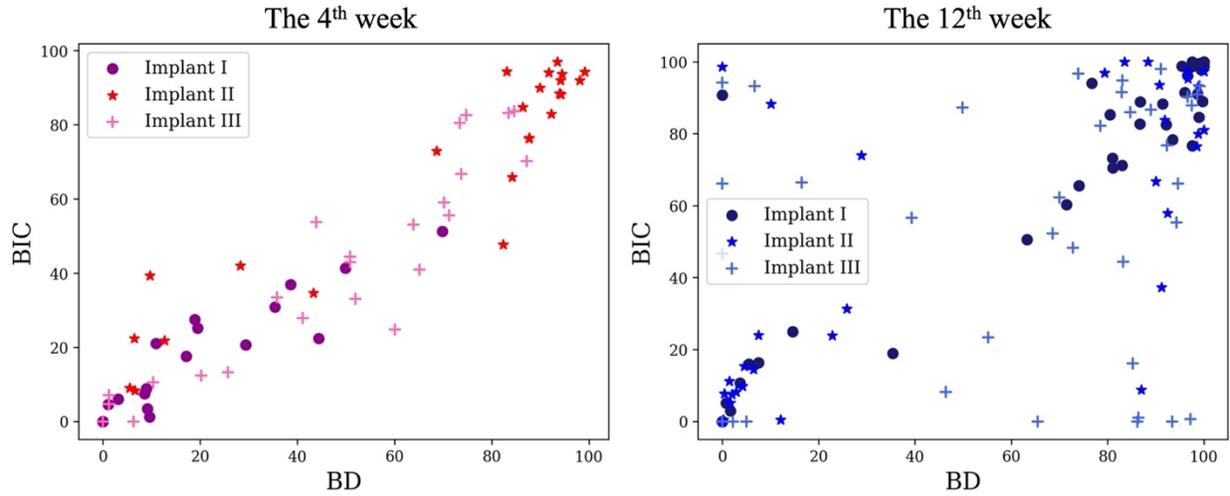

*Figure 4. Correlation between BD and BIC is positive at both week 4 (+0.96) and week 12 (+0.64).*

4. Discussion

Osseointegration is a bone healing process that reflects the establishment and maintenance of rigid fixation in bone subjected to functional loading[35]. This process occurs in several stages: initial stabilization of the implant under various stimuli, bone regeneration in peri-implant areas, and structural integration as a cohesive unit[36]. Because the early phase of osseointegration determines primary stability and eventually the success of fixation, the first 12 weeks after implantation are pivotal for evaluating bone ingrowth and remodeling. Although osseointegration is a complex process involving immune-inflammatory responses, angiogenesis, and osteogenesis[37], cellular activities and bone development can be effectively studied through in vivo experimental observations, particularly by using histological evaluations[38, 39].

In the present histological studies, the limitations of staining techniques and image magnification restricted this study's ability to analyze the quantity and location of osseous apposition with the implants. Empirically, a healing period of 4 to 12 weeks is sufficient to generate substantial amounts of mature lamellar bone and some immature woven bone[28]. Thus, differentiation into fibrous tissue, cartilage, or mature and immature bone types was not classified in this study. Instead, we focused on the volume of bone formation and BIC, which reflect the extent of bone ingrowth. Moreover, the physical and chemical characteristics of the implant surface, including roughness, topography, composition, energy and wettability[40–42], considerably affect cellular responses and events in peri-implant areas. Thus, we examined the bone-implant interface because it directly affects the progress and quality of osseointegration.

Our histological results reveal that Implant II (BG-coated) demonstrated superior bone regeneration to that of Implants I (pure Ti-6Al-4V) and III (BG-doped) at week 4, as indicated by higher BIC and BV values for Implant II. The ability of BGs to enhance bone formation has been demonstrated in many previous studies[17]. This effect is attributable to the release of ionic products, including apatite, calcium, and silicon ions, from BG[43]. These products can stimulate osteogenic activity by activating biological growth factors associated with bone formation. The dissolution of BG also facilitates the formation of a strong bond between the surrounding bone and the apatite layer on the implant surface. Thus, Implants II and III, both of which contained 58S, promoted rapid mineralization during the early stages of bone regeneration in this study.

In particular, BG-coated Implant II exhibited better bone ingrowth than did BG-doped Implant III. This difference can be attributed to the larger surface area of BG available in the coating compared with that achievable through the doping method. In addition, the rough surfaces produced by applying AM techniques, such as 3D printing, increase the surface area and enhance interlocking with the living bone[44,45]. The coated surface further promotes direct contact osteogenesis by supporting greater cellular proliferation and bone integration. However, the effects of BG addition may vary depending on factors such as the coating method, thickness, and composition. Future studies should compare the effects of BG addition through coating and doping methods.

The results of the present study indicate that after 12 weeks of healing, bone ingrowth reached similar levels among the three types of implants. However, pure Ti-6Al-4V (Implant I) exhibited higher BIC and BD values than did the two BG-added implants (Implants II and III). Moreover, numerous bubble-like white regions were observed in the bone (blue) surrounding Implants II and III, as indicated in Figure 2(h) and (i). These findings suggest that although BG demonstrated the ability to promote osteogenesis and angiogenesis through its high dissolution rate of ions that enhanced biological activities, particularly during the early postoperative stage[46], BG remains an imperfect material for supporting long-term bone growth. Van Dijk La et al. reported that the fusion rate of 45S5 BGs was inferior to that of autograft after 12 weeks of healing in an ovine posterolateral spinal model[47]. The calcium content in the BGs used in this study may have led to a high pH value. This phenomenon has been reported in the literature[48, 49] and occurs because of the rapid release of calcium, sodium, or other alkaline ions from BGs, leading to an environment that is unfavorable for cellular activity and detrimental to subsequent bone ingrowth. This likely explains why Implant II (BG-coated) exhibited the lowest average BD value after 12 weeks, whereas the impact on the BD of Implant III (BG-doped) was less pronounced. To address these challenges, pH-neutral BGs, as reported in the literature[50–52], can be used in future applications.

The phenomenon of "high BIC but low BD" and "high BD but low BIC" observed in Implant III suggests that the distribution of BG-doped particles in the additively manufactured implant is not uniform. This non-uniformity leads to inconsistent bone ingrowth performance around Implant III, as evidenced by the varying bone morphologies observed around its threads in

Figure 2(i). These bone morphologies include regions with no bone ingrowth (white), a thin layer of bone attached to the implant surface (high BIC but low BD), areas with a thin layer of unattached bone (high BD but low BIC), and threads fully filled with bone. This variability highlights the need for further research to improve the quality of additively manufactured BG-doped implants to achieve more uniform particle distribution and consistent osseointegration performance.

This study revealed the effect of BG addition during the early stages of bone healing. However, further more in-depth investigations are required. For example, the present study used only four rabbits in its animal tests; a larger sample size is required to achieve statistical significance and ensure generalizability for clinical applications. In addition, employing more advanced staining techniques in future studies could provide valuable insights into bone maturity and cell differentiation.. Future experiments could also include pH value measurements to optimize the composition of BGs. For additively manufactured implants, printing parameters are crucial because they affect surface conditions such as roughness and wettability. Thus, in addition to material composition, different surface conditions should be considered. Finally, extending the healing period in future studies would be advantageous for understanding the long-term remodeling process. According to previous studies on osseointegration with implants in rabbits[53,54], studies spanning 12 to 24 weeks could provide comprehensive insights into the progression of bone healing and integration.

5. Conclusion

Compared with implants made of pure Ti-6Al-4V, 3D-printed implants made of Ti-6Al-4V with 58S BGs added through either coating or doping are associated with enhanced bone ingrowth during the early stages of bone healing, as evidenced by the *in vivo* BD and BIC results. This finding is notable because rapid osseointegration is highly desirable in clinical applications. Furthermore, the current study demonstrated that BG can be effectively doped onto conventional 3D printing particles, such as Ti-6Al-4V, and that these doped particles can be successfully utilized in AM. Although BG-doped implants accelerate early-stage bone healing, their bone ingrowth performance remains inconsistent over the long term because of the non-uniform distribution of BG within the implant. Addressing challenges such as optimizing the pH balance of BG and achieving a consistent doping concentration is critical. If these problems are resolved, the proposed biomaterial and its applications hold considerable potential for becoming viable commercial implant materials in the future.

1. Furko M, Balázsi C. Calcium Phosphate Based Bioactive Ceramic Layers on Implant Materials Preparation, Properties, and Biological Performance. *Coatings*. 2020;10(9):823. https://doi.org/10.3390/coatings10090823
2. Li Y, Yang C, Zhao H, Qu S, Li X, Li Y. New Developments of Ti-Based Alloys for Biomedical Applications. *Materials*. 2014;7(3):1709–1800. https://doi.org/10.3390/ma7031709
3. Murr LE, Quinones SA, Gaytan SM, *et al.* Microstructure and mechanical behavior of Ti–6Al–4V produced by rapid-layer manufacturing, for biomedical applications. *Journal of the Mechanical Behavior of Biomedical Materials*. 2009;2(1):20–32. https://doi.org/10.1016/j.jmbbm.2008.05.004
4. Niinomi M. Low Modulus Titanium Alloys for Inhibiting Bone Atrophy. *Biomaterials Science and Engineering*. InTech; 2011 https://doi.org/10.5772/24549

5. Jiang X, Yao Y, Tang W, *et al.* Design of dental implants at materials level: An overview. *Journal of Biomedical Materials Research Part A*. 2020;108(8):1634–1661. https://doi.org/10.1002/jbm.a.36931

6. Niinomi M, Nakai M. Titanium-Based Biomaterials for Preventing Stress Shielding between Implant Devices and Bone. *International Journal of Biomaterials*. 2011;2011:e836587. https://doi.org/10.1155/2011/836587

7. Oh I-H, Nomura N, Masahashi N, Hanada S. Mechanical properties of porous titanium compacts prepared by powder sintering. *Scripta Materialia*. 2003;49(12):1197–1202. https://doi.org/10.1016/j.scriptamat.2003.08.018

8. Wu C, Luo Y, Cuniberti G, Xiao Y, Gelinsky M. Three-dimensional printing of hierarchical and tough mesoporous bioactive glass scaffolds with a controllable pore architecture, excellent mechanical strength and mineralization ability. *Acta Biomaterialia*. 2011;7(6):2644–2650. https://doi.org/10.1016/j.actbio.2011.03.009

9. Barrère F, van der Valk CM, Dalmeijer R a. J, van Blitterswijk CA, de Groot K, Layrolle P. In vitro and in vivo degradation of biomimetic octacalcium phosphate and carbonate apatite coatings on titanium implants. *Journal of Biomedical Materials Research Part A*. 2003;64A(2):378–387. https://doi.org/10.1002/jbm.a.10291

10. Hench LL, Splinter RJ, Allen WC, Greenlee TK. Bonding mechanisms at the interface of ceramic prosthetic materials. *Journal of Biomedical Materials Research*. 1971;5(6):117–141. https://doi.org/10.1002/jbm.820050611

11. Hench LL. Third-Generation Biomedical Materials. *Science*. 2002;295(5557):1014–1017. https://doi.org/10.1126/science.1067404

12. Rahaman MN, Day DE, Sonny Bal B, *et al.* Bioactive glass in tissue engineering. *Acta Biomaterialia*. 2011;7(6):2355–2373. https://doi.org/10.1016/j.actbio.2011.03.016

13. Polini A, Bai H, Tomsia AP. Dental applications of nanostructured bioactive glass and its composites. *Wiley Interdisciplinary Reviews: Nanomedicine and Nanobiotechnology*. 2013;5(4):399–410. https://doi.org/10.1002/wnan.1224

14. Boccardi E, Philippart A, Juhasz-Bortuzzo JA, *et al.* Uniform Surface Modification of 3D Bioglass®-Based Scaffolds with Mesoporous Silica Particles (MCM-41) for Enhancing Drug Delivery Capability. *Frontiers in Bioengineering and Biotechnology*. 2015;3. https://doi.org/10.3389/fbioe.2015.00177

15. Baino F, Fiume E, Barberi J, *et al.* Processing methods for making porous bioactive glass-based scaffolds—A state-of-the-art review. *International Journal of Applied Ceramic Technology*. 2019;16(5):1762–1796. https://doi.org/10.1111/ijac.13195

16. Tomsia A, Launey M, Lee J, Mankani M, Saiz E. Nanotechnology Approaches for Better Dental Implants. *The International Journal of Oral & Maxillofacial Implants*. 2011.

17. Jones JR. Review of bioactive glass: From Hench to hybrids. *Acta Biomaterialia*. 2013;9(1):4457–4486. https://doi.org/10.1016/j.actbio.2012.08.023

18. Tomsia AP, Lee JS, Wegst UGK, Saiz E. Nanotechnology for Dental Implants. *The International Journal of Oral & Maxillofacial Implants*. 2013;28(6):e535–e546. https://doi.org/10.11607/jomi.te34

19. Moritz N, Rossi S, Vedel E, *et al.* Implants coated with bioactive glass by $CO_2$-laser, an in vivo study. *Journal of Materials Science: Materials in Medicine*. 2004;15(7):795–802. https://doi.org/10.1023/B:JMSM.0000032820.50983.c1

20. Blaker JJ, Nazhat SN, Boccaccini AR. Development and characterisation of silver-doped bioactive glass-coated sutures for tissue engineering and wound healing applications. *Biomaterials*. 2004;25(7–8):1319–1329. https://doi.org/10.1016/j.biomaterials.2003.08.007

21. Oliver JN, Su Y, Lu X, Kuo P-H, Du J, Zhu D. Bioactive glass coatings on metallic implants for biomedical applications. *Bioactive Materials*. 2019;4(October 2019):261–270. https://doi.org/10.1016/j.bioactmat.2019.09.002

22. Jubhari EH, Dammar I, Launardo V, Goan Y. Implant coating materials to increase osseointegration of dental implant: A systematic review. *Systematic Reviews in Pharmacy*. 2020. https://doi.org/10.31838/srp.2020.12.6

23. Thomas MV, Puleo DA, Al-Sabbagh M. Bioactive Glass Three Decades On. *JLT*. 2005;15(6). https://doi.org/10.1615/JLongTermEffMedImplants.v15.i6.20

24. Zhong J, Greenspan DC. Processing and properties of sol-gel bioactive glasses. *Journal of Biomedical Materials Research*. 2000;53(6):694–701. https://doi.org/10.1002/1097-4636(2000)53:6<694::AID-JBM12>3.0.CO;2-6

25. Hamadouche M, Meunier A, Greenspan DC, *et al.* Bioactivity of Bioactive Sol-Gel Glasses Coated Alumina Implants. *Key Engineering Materials*. 2001;192–195:413–416. https://doi.org/10.4028/www.scientific.net/KEM.192-195.413

26. Fathi MH, Doostmohammadi A. Bioactive glass nanopowder and bioglass coating for biocompatibility improvement of metallic implant. *Journal of Materials Processing Technology*. 2009;209(3):1385–1391. https://doi.org/10.1016/j.jmatprotec.2008.03.051

27. Fathi MH, Doost Mohammadi A. Preparation and characterization of sol–gel bioactive glass coating for improvement of biocompatibility of human body implant. *Materials*

*Science and Engineering: A*. 2008;474(1–2):128–133. https://doi.org/10.1016/j.msea.2007.05.041

28. Lam T-N, Trinh M-G, Huang C-C, *et al.* Investigation of Bone Growth in Additive-Manufactured Pedicle Screw Implant by Using Ti-6Al-4V and Bioactive Glass Powder Composite. *International Journal of Molecular Sciences*. 2020;21(20):7438. https://doi.org/10.3390/ijms21207438

29. Chou Y-J, Hsiao C-W, Tsou N-T, Wu M-H, Shih S-J. Preparation and in Vitro Bioactivity of Micron-sized Bioactive Glass Particles Using Spray Drying Method. *Applied Sciences*. 2018;9(1):19. https://doi.org/10.3390/app9010019

30. Yuan L, Ding S, Wen C. Additive manufacturing technology for porous metal implant applications and triple minimal surface structures: A review. *Bioactive Materials*. 2019;4(1):56–70. https://doi.org/10.1016/j.bioactmat.2018.12.003

31. Gmeiner R, Deisinger U, Schönherr J, *et al.* Additive manufacturing of bioactive glasses and silicate bioceramics. *Journal of Ceramic Science and Technology*. 2015. https://doi.org/10.4416/JCST2015-00001

32. Wang H, Zhao B, Liu C, Wang C, Tan X, Hu M. A Comparison of Biocompatibility of a Titanium Alloy Fabricated by Electron Beam Melting and Selective Laser Melting. *PLOS ONE*. 2016;11(7):e0158513. https://doi.org/10.1371/journal.pone.0158513

33. Kilkenny C, Parsons N, Kadyszewski E, *et al.* Survey of the Quality of Experimental Design, Statistical Analysis and Reporting of Research Using Animals. *PLoS ONE*. 2009. https://doi.org/10.1371/journal.pone.0007824

34. Virtanen P, Gommers R, Oliphant TE, *et al.* SciPy 1.0: fundamental algorithms for scientific computing in Python. *Nature Methods*. 2020;17(3):261–272. https://doi.org/10.1038/s41592-019-0686-2

35. Dhinakarsamy V, Jayesh R. Osseointegration. *Journal of Pharmacy and Bioallied Sciences*. 2015;7(5):228. https://doi.org/10.4103/0975-7406.155917

36. Lacroix D, Prendergast PJ, Li G, Marsh D. Biomechanical model to simulate tissue differentiation and bone regeneration: Application to fracture healing. *Medical & Biological Engineering & Computing*. 2002;40(1):14–21. https://doi.org/10.1007/BF02347690

37. Feller L, Chandran R, Khammissa RAG, *et al.* Osseointegration: biological events in relation to characteristics of the implant surface. *SADJ : journal of the South African Dental Association = tydskrif van die Suid-Afrikaanse Tandheelkundige Vereniging*. 2014;69(3):112, 114–7.

38. Simion M, Benigni M, Al-Hezaimi K, Kim D. Early Bone Formation Adjacent to Oxidized and Machined Implant Surfaces: A Histologic Study. *International Journal of Periodontics and Restorative Dentistry*. 2015;35(1):9–17. https://doi.org/10.11607/prd.2217

39. Han J, Hong G, Lin H, *et al.* Biomechanical and histological evaluation of the osseointegration capacity of two types of zirconia implant. *International Journal of Nanomedicine*. 2016;Volume 11:6507–6516. https://doi.org/10.2147/IJN.S119519

40. Schwartz Z, Boyan BD. Underlying mechanisms at the bone–biomaterial interface. *Journal of Cellular Biochemistry*. 1994. https://doi.org/10.1002/jcb.240560310

41. Raines AL, Olivares-Navarrete R, Wieland M, Cochran DL, Schwartz Z, Boyan BD. Regulation of angiogenesis during osseointegration by titanium surface microstructure and

energy. *Biomaterials*. 2010;31(18):4909–4917. https://doi.org/10.1016/j.biomaterials.2010.02.071

42. Schwartz Z, Lohmann CH, Vocke AK, *et al.* Osteoblast response to titanium surface roughness and 1?,25-(OH)2D3 is mediated through the mitogen-activated protein kinase (MAPK) pathway. *Journal of Biomedical Materials Research*. 2001;56(3):417–426. https://doi.org/10.1002/1097-4636(20010905)56:3<417::AID-JBM1111>3.0.CO;2-K

43. Xynos ID, Edgar AJ, Buttery LDK, Hench LL, Polak JM. Ionic Products of Bioactive Glass Dissolution Increase Proliferation of Human Osteoblasts and Induce Insulin-like Growth Factor II mRNA Expression and Protein Synthesis. *Biochemical and Biophysical Research Communications*. 2000;276(2):461–465. https://doi.org/10.1006/bbrc.2000.3503

44. Gittens RA, Olivares-Navarrete R, Schwartz Z, Boyan BD. Implant osseointegration and the role of microroughness and nanostructures: Lessons for spine implants. *Acta Biomaterialia*. 2014;10(8):3363–3371. https://doi.org/10.1016/j.actbio.2014.03.037

45. Shaoki A, Xu J, Sun H, *et al.* Osseointegration of three-dimensional designed titanium implants manufactured by selective laser melting. *Biofabrication*. 2016;8(4):045014. https://doi.org/10.1088/1758-5090/8/4/045014

46. Hoppe A, Güldal NS, Boccaccini AR. A review of the biological response to ionic dissolution products from bioactive glasses and glass-ceramics. *Biomaterials*. 2011;32(11):2757–2774. https://doi.org/10.1016/j.biomaterials.2011.01.004

47. van Dijk LA, Barrère-de Groot F, Rosenberg AJWP, *et al.* MagnetOs, Vitoss, and Novabone in a Multi-endpoint Study of Posterolateral Fusion. *Clinical Spine Surgery*. 2020;1. https://doi.org/10.1097/BSD.0000000000000920

48. Schickle K, Zurlinden K, Bergmann C, *et al.* Synthesis of novel tricalcium phosphate-bioactive glass composite and functionalization with rhBMP-2. *Journal of Materials Science: Materials in Medicine*. 2011;22(4):763–771. https://doi.org/10.1007/s10856-011-4252-4

49. Schmitz SI, Widholz B, Essers C, *et al.* Superior biocompatibility and comparable osteoinductive properties: Sodium-reduced fluoride-containing bioactive glass belonging to the CaO–MgO–SiO2 system as a promising alternative to 45S5 bioactive glass. *Bioactive Materials*. 2020;5(1):55–65. https://doi.org/10.1016/j.bioactmat.2019.12.005

50. Ray S, Saha S, Rahaman SH, *et al.* An in vitro evaluation of the variation in surface characteristics of bioactive glass coated SS316L for load bearing application. *Surface and Coatings Technology*. 2019;377:124849. https://doi.org/10.1016/j.surfcoat.2019.08.003

51. Zhao H, Liang G, Liang W, *et al.* In vitro and in vivo evaluation of the pH-neutral bioactive glass as high performance bone grafts. *Materials Science and Engineering: C*. 2020;116:111249. https://doi.org/10.1016/j.msec.2020.111249

52. Zhang D, Hupa M, Hupa L. In situ pH within particle beds of bioactive glasses. *Acta Biomaterialia*. 2008;4(5):1498–1505. https://doi.org/10.1016/j.actbio.2008.04.007

53. Borrajo JP, Serra J, González P, León B, Muñoz FM, López M. In vivo evaluation of titanium implants coated with bioactive glass by pulsed laser deposition. *Journal of Materials Science: Materials in Medicine*. 2007;18(12):2371–2376. https://doi.org/10.1007/s10856-007-3153-z

54. Hamadouche M, Meunier A, Greenspan DC, *et al.* Long-termin vivo bioactivity and degradability of bulk sol-gel bioactive glasses. *Journal of Biomedical Materials Research*.

2001;54(4):560–566. https://doi.org/10.1002/1097-4636(20010315)54:4<560::AID-JBM130>3.0.CO;2-J